# Transport characterization and quantum dot coupling in commercial 22FDX®

G.A. Elbaz, P.-L. Julliard, M. Cassé, H. Niebojewski, B. Bertrand, G. Roussely, V. Labracherie, M. Vinet, T. Meunier, B. Cardoso Paz

*Abstract*—Different groups worldwide have been working with the GlobalFoundries[TM] 22nm platform (22FDX®) with the hopes of industrializing the fabrication of Si spin qubits. To guide this effort, we have performed a systematic study of six of the foundry's processes of reference (POR). Using effective mobility as a figure of merit, we study the impact of gate stack, channel type and back bias as a function of temperature. This screening process selected qubit devices that allowed us to couple quantum dots along both the length and width of the Si channel. We present stability diagrams with clear and regular honeycomb patterns, where spurious elements such as dopants are not observed. By combining these results with room and low temperature simulations, we provide insights into potential technology optimizations and show both the utility of qubit pre-screening protocols as well as the advantages of leveraging forward body bias within an FDSOI (Fully Depleted Silicon-On-Insulator) qubit platform.

## I. Introduction

SIGNIFICANT advances have been made in the field of quantum computing, with the largest test platforms now surpassing a thousand physical qubits [1]. However, many current experimental platforms for quantum computing are based on technologies that suffer from poor scalability. Leveraging CMOS manufacturing would enable the rapid scale-up of quantum computing hardware, increasing both the number of qubits per square mm and the number chips produced. As such, there has been considerable effort in the last 12 years [2] to develop the Si spin qubit and bring it to the 300mm scale. Among the commonly explored Si-based technologies, Fully Depleted Silicon-On-Insulator (FDSOI) has proven to be a promising forerunner [3]. FDSOI has a unique local back gate which offers very strong control over the transistor threshold voltage ($V_{TH}$). In qubits, this back gate can also control the vertical position along the Si channel in which a quantum dot (QD) is formed, drawing it away from inherent defects at the front-gate Si/SiO$_2$ interface and reducing the charge-noise associated with interface traps [4].

While some progress has been made to bring FDSOI spin qubits to the commercial scale [5-8], the deep cryogenic temperatures and long measurement protocols necessary to fully characterize qubits can be prohibitive in determining the efficacy and variability of a new process split. Rigorous characterization routines must be developed to screen potential qubit technologies and aid process development. Effective electron mobility is sensitive to many variables within the process flow and can shed light on both material and interface defects [9]. Detailed information on these scatterers can be found by measuring the effective mobility as a function of temperature and carrier concentration. This feedback can then be leveraged to select the best process splits and target the die and quantum devices most likely to succeed for advanced testing.

In this work, we present a detailed study on the effective mobility of six process splits from the GlobalFoundries[TM] 22nm platform (22FDX®), varying channel type, front-gate stack and front-gate equivalent oxide thickness (EOT). We use both temperature and back gate bias along with TCAD simulations to inform our understanding of the process splits and choose the best split and back bias conditions for forming QDs within the integrated qubit structures. We show clear evidence of electrostatically controlled QDs within short qubit arrays (1x3 and 2x3) as well as capacitive tunnel coupling between said QDs in the x- and y- directions. Together, our data supports the selection of a thicker gate oxide in NMOS qubit arrays, which displays improved QD coupling relative to what has been reported for thinner gate oxides in 22nm FDSOI. [8]

## II. DEVICES AND METHODOLOGY

All devices were fabricated using the commercially available 22FDX® platform from GlobalFoundries[TM] with their process of reference (POR). The buried oxide and silicon channel in this technology are 20nm and 6nm thick, respectively. The commissioned wafers were designed with a wide range of in-wafer and in-die process splits as well as integrated test structures which allowed us to screen several potential front and

This work was supported in by the European Research Council (ERC) Synergy QuCube (grant number 810504), European Innovation Council (EIC) MCSQUARE (grant number 101136414), and the French National Research Agency under the program "France 2030" (PEPR PRESQUILE - ANR-22-PETQ-0002).

G. A. Elbaz, P.-L. Julliard, M. Vinet, T. Meunier and B. Cardoso Paz are with Quobly, Grenoble, FR 38040 (e-mail: bruna.cardoso-paz@quobly.io).
M. Cassé, H. Niebojewski, B. Bertrand, G. Roussely and V. Labracherie are with the Université Grenoble Alpes, CEA-LETI, Grenoble, FR 38040.

back gate stacks. The process splits discussed in this study are outlined in Table I. The first is focused on gate stack materials (GS-1 *vs*. GS-2), the second between thin (EOT-1 = 1.25 nm) and thick (EOT-2 = 3.4 nm) gate oxides. These were then tested on both n-type and p-type Si channels.

TABLE I
IN DIE PROCESS SPLITS

| Split | Gate Stack (GS) | Oxide (EOT) | Carrier type |
|---|---|---|---|
| A | GS-1 | Thin (EOT-1) | PMOS |
| B | GS-2 | Thin (EOT-1) | |
| C | GS-2 | Thick (EOT-2) | |
| D | GS-1 | Thin (EOT-1) | NMOS |
| E | GS-2 | Thin (EOT-1) | |
| F | GS-2 | Thick (EOT-2) | |

Test structures were transistors, 9μm x 10μm in size, with a local back gate. Two styles of qubit devices were measured in this study: one which contains several gates in series (SR), with a similar design as in [10], and a second, which contains a 2xN array of split gates in series [11], often called face-to-face (FF).

Variable temperature I-V and C-V measurements were collected from 300K down to 4.2K in a Lakeshore Probe Station, using a commercial Semiconductor Device Parameter Analyzer and LCR-meter. Wafer-level room temperature data were collected on a Cascade Semi-automatic Probe Station. QD experiments were carried out in a dilution refrigerator at 100mK. Simulations were carried out by Sentaurus TCAD from Synopsys Inc. Metal gate work functions and gate oxide dielectric constants were calibrated on experimental capacitance measurements. Density correction based on a quantum correction model was employed with the values calibrated in [12].

## III. QUBIT PROCESS-SPLIT PRE-SCREENING

In this section we discuss in detail the variable-temperature mobility of 22nm FDSOI NMOS and PMOS transistors as a function of back bias, gate stack and equivalent oxide thickness (EOT). This discussion is complemented by TCAD simulations, which support the choice for EOT-2 and forward body biasing in 22FDX® as well as shed light into some of the scattering and transport mechanisms at play in MOSFETs and qubits alike at cryogenic temperatures.

### A. Split C-V Mobility

Carrier mobility is an important indicator for high-quality qubits. We therefore used split C-V to screen the six potential process splits by studying the behavior of 22FDX® MOSFETs with lowering temperatures and back bias. Figure 1 plots the effective carrier mobility, $\mu_{eff}$ as a function of temperature at zero body bias (ZBB, $V_{BG}$ = 0). As expected, the mobility of each device increases with decreasing temperature. However, some differences between PMOS and NMOS emerge. If we first compare gate stacks, this has a negligible effect on NMOS, while in PMOS the first gate stack (GS-1), consistently shows ~50% lower mobility than GS-2. If instead we compare the effects of equivalent oxide thickness, NMOS proves itself to be more sensitive. While PMOS EOT-2 does benefit from slight improvements in mobility (ex. – 334 vs. 401 cm$^2$/Vs maximum at 4.2K), NMOS mobilities for EOT-2 are significantly larger than EOT-1 at every temperature, increasing from 1.7x to 3.6x greater at 300K and 4.2K respectively. This is in line with published results showing that mobility reduces with decreasing EOT for high-*k* gate oxides due to increased optical phonon scattering, whereas the quality of the high-*k* gate oxide determines the amount of remote charge scattering. [13]

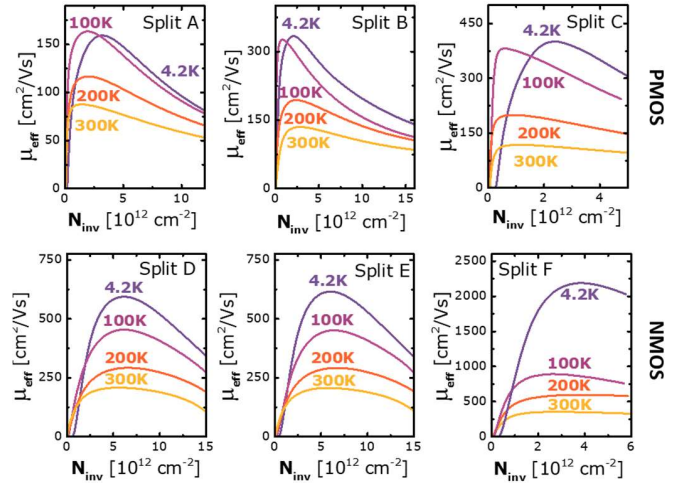

Fig. 1. Effective carrier mobility ($\mu_{eff}$) *vs*. the inversion charge density ($N_{inv}$) as a function of temperature for both front gate stacks and equivalent oxide thicknesses when $V_{BG}$ = 0V (ZBB).

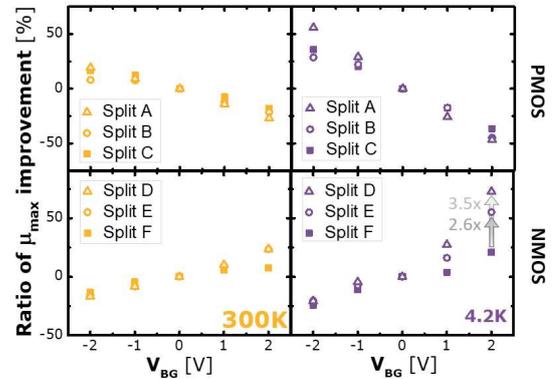

Fig. 2. Ratio of maximum effective mobility ($\mu_{max}$) improvement w.r.t. $V_{BG}$ = 0 vs. the back gate voltage ($V_{BG}$) at (left) 300K and at (right) 4.2K for the six process splits: GS-1 (triangle) and GS-2 (circle, square), EOT-1 (open markers) and EOT-2 (closed markers).

Irrespective of temperature, the back gate has a significant impact on the maximum $\mu_{eff}$ ($\mu_{max}$). In Fig. 2 we plot the relative change in $\mu_{max}$ with applied back bias of ±1, ±2V as compared to ZBB and look at how this difference is modified with temperature. To consistently compare NMOS and PMOS, we refer to these values as forward body biasing (FBB) and reverse body biasing (RBB), as the back bias polarity is inverted for the two channel types. In all cases, FBB improves the $\mu_{max}$, while reverse RBB decreases it.

If we first compare GS-1 and GS-2, at 4.2K, where the $\mu_{eff}$ is mainly limited by the channel-oxide interface surface roughness, we see stronger improvements under FBB in GS-1 (Split A and D) than GS-2. This difference is larger in NMOS, where we see a 73% increase in $\mu_{max}$ with a FBB = 2V for GS-1 in Split D *vs*. 55% and 21% for GS-2 (Splits E and F, respectively). The relatively low $\mu_{max}$ improvement for Split F can be explained, in part, by the fact that the ZBB mobility in EOT-2 is closer to the universal mobility and so there is less ground to be gained with back biasing. That said, with this 21% increase in mobility at lower charge carrier densities, EOT-2 reaches a $\mu_{max}$ of 2645cm$^2$/Vs under FBB. This maximum is on par with comparable FDSOI transistors with thicker gate oxides (EOT = 6nm) and no high-*k* materials [14]. This is significant because high-*k* materials are a known source of defects which degrade mobility by introducing more scattering centers when compared to SiO$_2$.

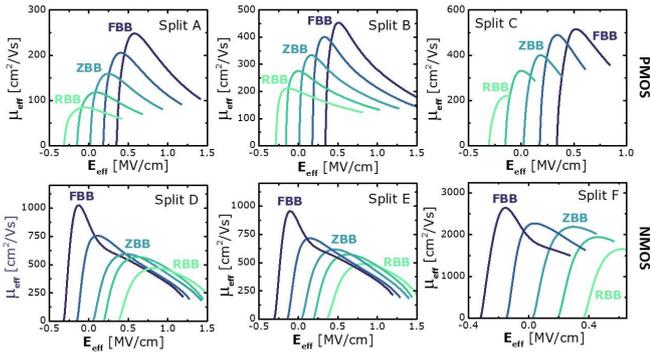

Fig. 3. Effective carrier mobility ($\mu_{eff}$) *vs.* the effective electric field ($E_{eff}$), as a function of applied back gate voltage ($V_{BG}$) at intervals of 1V (max = ±2V) for both front gate stacks and equivalent oxide thicknesses at 4.2K.

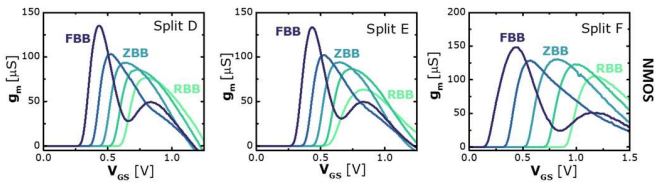

Fig. 4. NMOS conductance ($g_m$) at 4.2K as a function of applied back gate voltage ($V_{BG}$) at intervals of 1V (max = ±2V) for both front gate stacks and and equivalent oxide thicknesses at 4.2K. A double peak appears at FBB = 2V.

To explore this, we take a closer look at the back-bias dependence at 4.2K. Figure 3 plots the split C-V $\mu_{eff}$ as a function of effective electric field, $E_{eff}$ [15], and $V_{BG}$. The shape of the effective mobility for NMOS devices is quite different than those of PMOS. At $E_{eff} \approx 0$ MV/cm to 0.2 MV/cm, the mobility for NMOS under a FBB = 2V displays a change in slope, which is indicative of two-channel conduction and suggests the presence of intersubband scattering (ISS) [16, 17]. This is supported by the NMOS transconductances ($g_m$), plotted in Fig. 4 as a function of $V_{GS}$, each of which shows two peaks. For both equivalent oxide thicknesses, the second peak lines up with RBB = 2V, which should exhibit strictly front-channel conduction. It is worth noting that evidence of ISS is not observed in the equivalent PMOS devices. ISS is generally observed when carrier mobility is large enough to be affected by this scattering mechanism; holes however have a larger effective mass which in our devices degrades their mobility beyond this point.

### B. Conduction Channels in EOT-1 vs. EOT-2 Devices

To confirm the conductance behavior and presence of two-channel conduction in the NMOS devices, we performed room and low temperature TCAD simulations on NMOS EOT-1 and EOT-2. Fig. 5 and 6 show the low temperature TCAD simulations of EOT-1 and EOT-2 of the relative electron density ($n_e$) as a function of position within the height of the channel ranging from a depth of 0 nm (front gate) to 6 nm (back gate). In Fig. 5 we show the low temperature evolution of this density as a function of applied $V_{GS}$, starting from $V_{GT} = 0V$ and increasing to $V_{GT} = 1V$ while in Fig. 6a, we focus on $\mu_{max}$ and show the densities calculated at the applied-$V_{GS}$ where $\mu_{max}$ occurs in the two devices at room and low temperature. In Fig. 6b we plot the mean location of the normalized electronic density within the channel ($\bar{z}$) at low temperature as a function of $V_{GS}$.

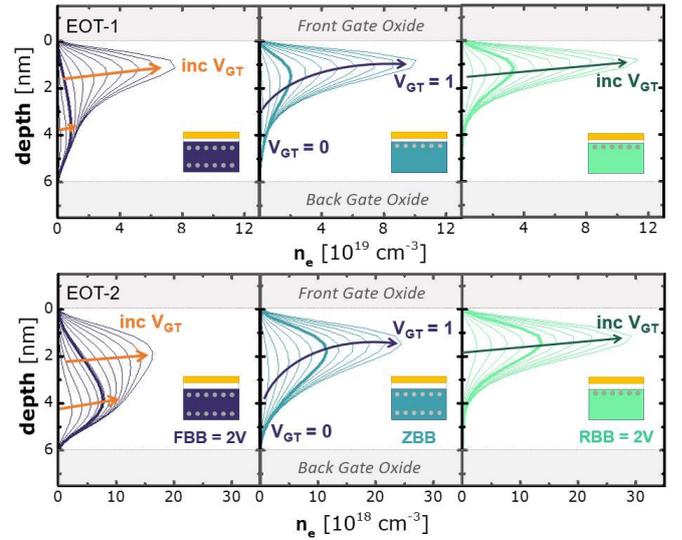

Fig. 5. Low temperature simulation of electron density as a function of depth within the Si channel for NMOS EOT-1 (top) and EOT-2 (bottom) with $V_{GT} = V_{GS} - V_{TH}$ ranging 0V to 1V (intervals of 100mV). $V_{GT}$ corresponding to $\mu_{max}$ is shown in bold. Results are plotted for three $V_{BG}$ values (0V, ±2V) spanning forward, zero and reverse back bias at 20K.

As we have previously reported for 28FDSOI with gates oxides similar to EOT-1, in simulations of 22FDSOI we found that carrier density distributions along the height of the Si channel at low temperatures are more localized than those at 300K, irrespective of back-gate voltage [18]. That said, at ZBB, in Fig. 6a we observe a more even distribution within the channel for EOT-2; at $\mu_{max}$ EOT-2 displays 9% and 5.5% more electron density than EOT-1 in the bottom half of the channel at room and low temperature, respectively (see Table II). In agreement with experimental results, this percentage decreases significantly for RBB ($V_{BG}$ = -2V), where 88-95% of the electron density is centered in the first 3nm of the channel in both device types at $\mu_{max}$. In comparison, more than half of the electron density can be found at the back interface (> 3nm) at $\mu_{max}$ for FBB ($V_{BG}$ = 2V) at both 300K and low temperatures for both EOTs.

A close look at the evolution of e$^-$ density with $V_{GT}$ under FBB further shows that the local minima in conductance ($g_m$) seen in Fig. 4 for both EOT-1 and EOT-2 occurs at $V_{GT}$ where ~50% of e$^-$ density can be found in the front and back halves of the channel. In other words, the largest intersubband scattering in these devices occurs at approximately equal population of the two channels. Furthermore, in Fig. 5 we see that EOT-2 maintains a large e$^-$ density at the back interface at high inversion, where $V_{GT}$ is well past $\mu_{max}$. In fact, 40% of the e- density can be found at $z > 3$nm when $V_{GT} = 1$. This is in contrast to EOT-1 which only has 16% at its back interface at $V_{GT} = 1$. As seen in Fig. 6b, the average e- position, $\bar{z}$, for EOT-2 does not decrease below 3nm and move towards the front gate conduction channel until $V_{GS} > 0.9$V (*i.e.* $V_{GT} > 0.7$V).

TABLE II
PERCENT ELECTRON DENSITY LOCATED IN THE
BOTTOM HALF OF THE SI CHANNEL (3 > z > 6) AT $\mu_{MAX}$

| T | EOT | FBB = 2V | ZBB | RBB = 2V |
|---|-----|----------|-----|----------|
| 300K | EOT-1 | 50.2 % | 20.8 % | 8.3 % |
|      | EOT-2 | 58.0 % | 30.8 % | 12 % |
| low T | EOT-1 | 56.5 % | 16.5 % | 4.7 % |
|       | EOT-2 | 60.4 % | 22.2 % | 7.0 % |

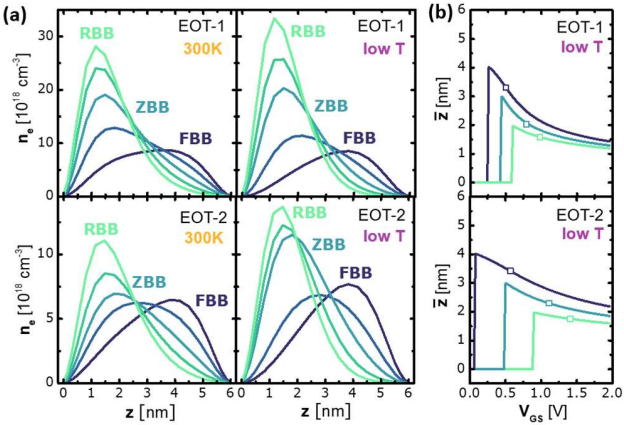

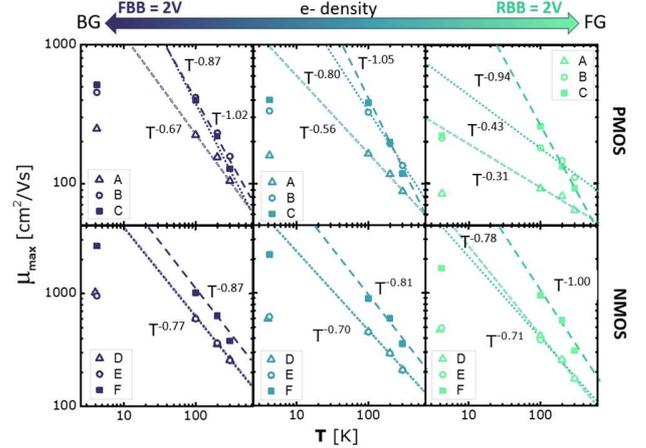

Fig. 6. (a) TCAD simulations of electron density ($n_e$) *vs*. Si channel depth ($z$) at $\mu_{max}$ for EOT-1 and EOT-2 at 300K and low temperature prior to any ISS as a function of applied back bias (intervals of 1V). (b) mean location of $n_e$ ($\bar{z}$) as a function of $V_{GS}$ for forward (dark purple), zero (blue) and reverse (light green) body bias ($V_{BG} = 0, \pm 2V$). Open square marker corresponds to $\mu_{max}$.

Fig. 7. Maximum effective carrier mobility ($\mu_{max}$) *vs*. temperature on log-log scale for GS-1 (triangle) and GS-2 (circle, square), EOT-1 (open markers) and EOT-2 (closed markers) for three $V_{BG}$ values (0V, ±2V). A power law with temperature is shown for each case.

To fully understand the implications of these results, we return to the experimental data in Fig. 7 and plot $\mu_{max}$ *vs*. temperature on a log-log scale as a function of gate stack and back gate bias ($V_{BG}$). All devices above 100K display a power law $T^{-\gamma}$, which has been shown to be primarily correlated with the mechanism for phonon scattering ($\mu_{ph}$) within the device [19]. While different bulk MOSFET technologies have reported $\gamma > 1$, with differences correlated to the channel thickness [20], $0.7 \leq \gamma \leq 1$ is typically observed for FDSOI devices [21]. As can be seen in Fig.7, our devices display a $\gamma$-dependence on gate stack, EOT and $V_{BG}$, suggesting that these all affect the scattering mechanisms within the Si channel. As with mobility at 4.2K, the PMOS temperature dependence with $V_{BG}$ is more strongly affected by gate stack materials while with NMOS, the EOT has a larger role to play.

Combining these results with our simulations of e-density location within the Si channel, the differences observed in $\gamma$ with back bias paint a clear picture. In EOT-1, the gate oxide is thin and the carriers thus experience more Coulomb scattering from the high-*k* above when traveling near the top surface, as occurs with ZBB and RBB. This effect is compounded in PMOS by the surface roughness of GS-1 in split A, (so much so that the mobility *decreases* at 4.2K w.r.t. 100K). The effect of this scattering is reduced under FBB, as the carriers are pulled down towards the back gate and $\mu_{max}$ corresponds to an e- distribution primarily in the middle of the Si channel, particularly at warmer temperatures, leading to an increase in $\gamma$. In this regime, the quality of the BOX interface starts to have a stronger influence. In comparison, the thicker oxide of EOT-2 shifts the baseline conduction pathway of the carriers such that, under both ZBB and FBB, most of the carriers are found in the middle and back half of the channel respectively when $\mu_{max}$ occurs, leading to $\gamma \sim 0.9$ in EOT-2 under FBB. However, under RBB, while most of the carriers are again drawn to the front interface for all $V_{GS}$, EOT-2 has much less Coulomb scattering, as the gate oxide used is thick enough to shield the carriers from the high-*k*, leading to an observed $\gamma \sim 1$ for both carrier types.

Meanwhile, at carrier densities higher than that of $\mu_{max}$, the oscillations observed in electron carrier mobility under FBB along with the strong evidence for two-channel conduction suggest that the electron mobility in both EOT-1 and EOT-2 is degraded by ISS when operated at high inversion densities. The mobility improvements observed under FBB could therefore be limited by the interaction between front and back channels in ultrathin silicon films. This interaction, and subsequent ISS, may be ameliorated in thicker Si films, where the two conduction channels would be better physically separated, and should be considered for future quantum technologies.

## IV. SPLIT F QUBIT LINEAR AND BILINEAR ARRAYS

With the insights from the in-depth characterization of our process splits found in Section III, we selected Split F (NMOS, GS-2, EOT-2) operated under FBB as the best candidate for qubits. In this section we present our findings on qubit arrays fabricated with this split. We share both MOSFET characteristics at room and low temperature as well as single-electron transistor (SET) characteristics. Finally, as a coupled double-QD is the minimum cell unit necessary to demonstrate a Si spin two-qubit gate, we demonstrate clear capacitive and tunnel coupling between two quantum dots within our qubit arrays and present honeycomb patterns for each device type that show no spurious elements.

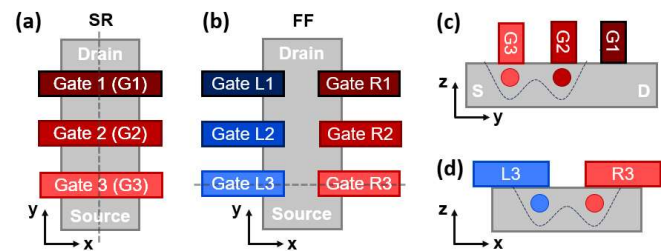

Fig. 8. (a) – (b) Top-down schematic of our qubit designs consisting of (a) three gates in series (SR device) and (b) three face-to-face or split gates (FF device). Gate pitch: 90nm, width: (a) 40nm (b) 70nm. (c) – (d) Cross-section schematic of two adjacent QDs coupled in (c) an SR device, along the y-axis and (d) an FF device, along the x-axis.

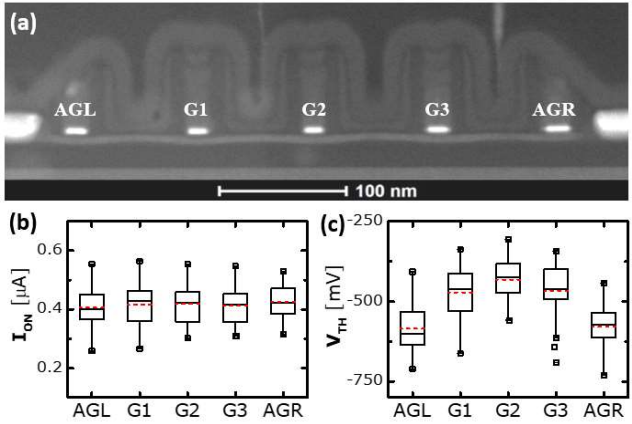

Fig. 9. (a) TEM of a process split F 1x3 SR qubit array (b)-(c) box and whisker plot of (b) $I_{ON}$ at $V_{GT}$ = 0.5 V (c) $V_{TH}$ as a function of qubit gate for process split F under FBB = 3V. Red dashed line represents the mean over ~40 dies.

Schematics of our qubit devices can be found in Fig. 8a and 8b, respectively. In SR devices (1xN arrays), QDs are formed under the individual gates and are capacitively coupled along the direction of the Si channel length (y-axis), as shown in Fig. 8c. In FF devices (2xN arrays), the capacitive coupling is opened to two dimensions and lateral coupling across the width of the Si channel (x-axis) is now feasible (Fig. 8d). The two-dimensional nature of this coupling makes our arrays compatible with the two most prevalent readout techniques for Si spin qubits today: RF-reflectometry [10] and nearby SET charge detection [22].

### A. Split F MOSFET characteristics

As Si-spin qubits are modified MOSFETs, we first characterize their behavior in the transistor-like (FET) regime prior to the single-electron transistor (SET) regime. In Fig. 9 we show a TEM of an SR qubit (1x3 array with 2 access gates AGL and AGR) qubit along with the distribution of room temperature on-state currents ($I_{ON}$) and threshold voltages ($V_{TH}$) collected for each individual gate (with all other gates operated in the strong inversion regime), under FBB = 3V, across a full wafer (~40 dies distributed over center, middle and edge radii). $I_{ON}$ at $V_{GT}$ = 0.5V shows similar metrics for all 5 gates, with an average value of ~425nA. As expected, the outermost gates (AGL and AGR) show a more negative $V_{TH}$ because of their proximity to the highly doped source-drain electrodes; diffusion of dopants from this region create a non-uniform electrostatic environment. Systemic shifts in $V_{TH}$ such as these have been observed in other Si qubit technologies [23]. We observe approximately a 100mV range between the upper and lower quartiles and a 300mV range total, which is also on par with other Si qubit technologies. While this range is respectable, to further minimize within wafer effects, the QDs in this study were selected from the same die as the test structures in Section III.

To operate the qubit device as a standard transistor at cryogenic temperatures, we shorted all gates (AGL, G1 – G3, AGR) and collected the current that flows from drain to source ($I_{DS}$) as a function of the front-gate voltage ($V_{GS}$) with drain bias ($V_{DS}$) of 50mV. Fig. 10a shows $I_{DS}$ normalized by the channel width ($I_{DS}$/W) measured at different FBBs ($V_{BG}$ = 0 to 2V). These display current oscillations due to the formation of subbands at cryogenic temperatures and are indicative of quantum confinement [24]. Fig.

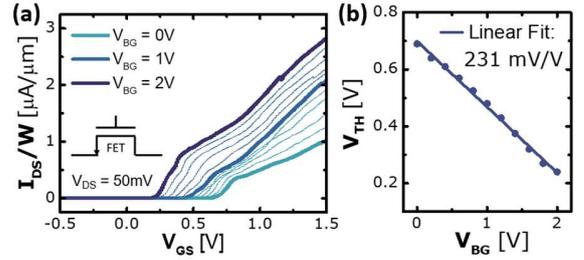

Fig. 10. Process split F 1x3 SR qubit array operated in the FET regime ($V_{DS}$ = 50 mV): (a) $I_{DS}$/W *vs*. $V_{GS}$ sweeping all top gates simultaneously as a function of applied back bias ($V_{BG}$) under FBB (intervals of 0.2V) (b) Threshold voltage, $V_{TH}$ (circles) *vs*. $V_{BG}$ from $I_{DS}$-$V_{GS}$ in (a). Linear fit (solid line) has $R^2 \geq 0.99$.

10b shows the extracted $V_{TH}$ *vs*. $V_{BG}$. From the linear dependence of $V_{TH}$ on $V_{BG}$, we find a body factor, $\Delta V_{TH}/\Delta V_{BG}$, of 231mV/V. The body factor is greater than that of the split F MOSFETs studied in Section III ($\Delta V_{TH}/\Delta V_{BG}$ = 188mV/V); the gaps between neighboring front gates give the back gate an even stronger capacitive control with respect to the front gates.

### B. SR devices: QDs coupled along the y-axis

Operation in the SET regime requires a $V_{DS}$ significantly smaller than the energy separation between subbands. The spaces between gates (defined by the gate pitch) then act as barriers and promote 3D confinement. We therefore used $V_{DS}$ = 1mV and swept individual gates while operating the other two gates in series in the strong inversion regime ($V_{GS} \gg V_{TH}$). This configuration forms a single QD under the swept gate, while the other gates operate as extensions of the "reservoirs" (*i.e.*, source and drain). Fig. 11a shows the results from one such set of sweeps as a function of G3 at $V_{BG}$ = 2V. The resulting $I_{DS}$-$V_{GS}$ displays many well-defined peaks, known as Coulomb peaks, with each peak corresponding to the overall addition of a single electron to the QD [25,26]. On a given peak, the associated chemical potential of the QD is in resonance with the reservoirs and electron-electron repulsion prevents more than a single electron from tunneling into and out of the QD at a time. The presence of many equally spaced peaks indicates that a true QD has been created within our qubit device and the signal is not due to a dopant or other defect displaying quantum confinement-like behavior.

Next, we repeat this experiment for a range of $V_{DS}$, and present the data in Fig. 11b. As expected for a QD SET, a diamond-like pattern is observed, (referred to as Coulomb diamonds), where a given diamond represents the $V_{DS}$ and $V_{GS}$ needed to maintain a stable N-electron configuration; within a diamond, no additional electrons can tunnel into the QD due to Columb blockade [27,28]. Successive diamonds represent states with N+1, N+2, etc. electron counts. From the heights and widths of the Coulomb diamonds, we can extract two important parameters for the QD: the charging energy, $E_C$, and the lever arm, α. The charging energy ($E_C$) represents the energy required to add or remove an electron to the QD and is inversely proportional to the radius of the QD [27]. For G3, we calculate a range of charging energies, with $E_C$ varying between 2.0 and 4.5meV. This range stems from the irregularity of the Coulomb diamond heights and reveals some variability in the size of the QD. The source of this variability is likely due to a localized defect which is exerting unwanted electrostatic force on the QD. The lever arm (α) is a measure of the electrostatic capacitive control the gate exerts on the QD and was found to range from 0.2 and 0.4 eV/V for the many Coulomb diamonds in Fig. 11b, which is on par with other CMOS-based QDs [22].

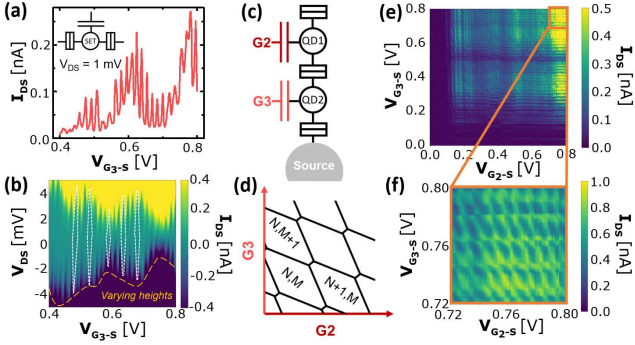
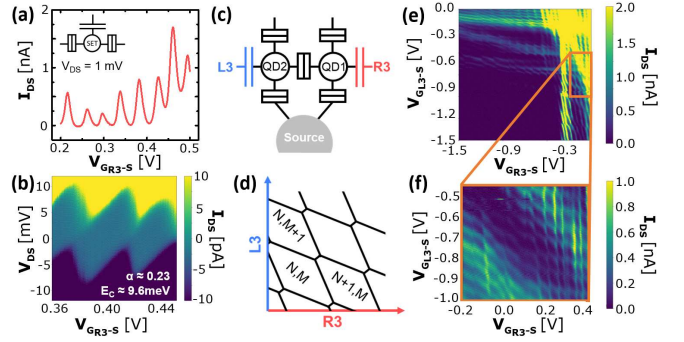

Fig. 11. Process split F 1x3 (SR) qubit device operated under FBB in (a) – (b) the SET regime displaying characteristic (a) Coulomb peaks (b) Coulomb diamonds while sweeping G3. Several Coulomb diamonds are outlined in white to guide the eye. A difference in heights is observed, leading to a range of charging energies and lever arms. (c) Electronic schematic of two adjacent QDs coupled in an SR qubit device, along the length of the Si channel (y-direction). (d) Theoretical stability diagram with the honeycomb pattern predicted for two coupled QDs. (e) and (f) Stability diagrams of two coupled quantum dots under FBB, located underneath gates G2 and G3, in an SR qubit device.

Fig. 12. Process split F 2x3 (FF) qubit device operated under FBB in (a) – (b) the SET regime displaying characteristic (a) Coulomb peaks (b) Coulomb diamonds while sweeping R3, with charging energies of ~9.6meV and a lever arm of ~0.23 eV/V. (c) Electronic schematic of two adjacent QDs coupled along the x-direction in an FF qubit device. (d) Theoretical stability diagram with the honeycomb pattern predicted for two laterally coupled QDs. (e) and (f) Stability diagrams of two coupled QDs under FBB, located underneath gates L3 and R3, in an FF qubit device.

To couple multiple QDs, we repeat the $I_{DS}$-$V_{GS}$ under FBB (2V) at very low $V_{DS}$ (1 mV) but as a function of two front gates rather than one (with the last qubit and both access gates still operated in the strong inversion regime.) In this case, two QDs are formed, one under each swept gate, as shown schematically in Fig. 11c. For two quantum dots capacitively and tunnel coupled, the 2D $I_{DS}$-$V_{GS}$ map of the two swept gates (*i.e.*, the stability diagram) should display a "honeycomb" pattern as depicted in Fig. 11d, which represents the equilibrium charge states of the coupled QDs. This pattern arises from the crossing of Coulomb peaks from each QD, where, for the example shown in Fig. 11d, each successive horizontal and vertical line corresponds to an additional electron added to the G2 and G3 QDs respectively. An angled slope of the vertical and horizontal lines indicates that the dots are capacitively coupled [29].

In Fig. 11e and 11f we show this stability diagram for our SR qubit device with 2 QDs formed under gates G2 and G3. In the many-electron regime, we see a clear honeycomb pattern in Fig. 11f, indicating that the two adjacent QDs are indeed capacitively and tunnel coupled. In our case, we observe many peaks within the stability diagram, and we estimate that our SR qubit structure can hold a minimum of 35-40 electrons per quantum well before a MOSFET-like behavior is observed. The barrier between the two QDs defines their coupling and is primarily determined by the gate pitch and $V_{BG}$. A gate spacing of 60nm is therefore large enough to form barriers within 22FDX® and define the quantum wells. To further fine tune this coupling and have individual control over this barrier, a second level of gates could be integrated [30]. Alternatively, as demonstrated in [7], control over the barrier can be reached in linear arrays if the QDs are formed between the gates.

### C. FF devices: QDs coupled along the x-axis

Many Si spin qubit devices rely on having a 2xN array-like structure such that one side of the array can be used to sense the other and read-out its state. Moreover, in this type of array, each qubit can have more closest "neighbors" (*i.e.*, more entangled qubits). Therefore, it is also important to be able to couple QDs in the *x*-direction. We use our FF qubit device to probe this functionality in 22FDX®. As in the SR device, we observe many characteristic Coulomb peaks (Fig. 12a) and Coulomb diamonds (Fig. 12b) under FBB, shown here for the bottom-right gate, R3. From the Coulomb diamonds we extract charging energy of 9.6meV and a lever arm of 0.23 eV/V. The larger $E_C$ measured for the QD under gate R3 in this device implies that this quantum dot is smaller than the one measured in the SR qubit device. This is consistent with the difference in the gate areas between the two device styles, which changes from 1200nm² to 300nm² for SR and FF, respectively. The modification of the QD size with changing gate size is a good confirmation that the QDs that we are measuring are electrostatically defined underneath the front gates as desired and that the behaviors observed are not due to defects which can often display similar Coulomb blockade signatures [31]. The lever arm, by comparison, is on par with that of the SR qubit and other CMOS-based QDs [22].

To couple QDs in a face-to-face configuration, we again repeat the $I_{DS}$-$V_{GS}$ under FBB with very low $V_{DS}$ but as a function of two parallel front gates (with the other four qubit gates and access gates operated in the strong inversion regime.) The resulting stability diagrams can be seen in Fig. 12e and 12f, where the quintessential angled honeycomb pattern is again observed, this time for the bottom left and right gates (L3 and R3).

Recent work in 22FDX® has been published on similar qubit layouts with the same reported gate stack as Split E: GS-2 and EOT-1. Under similar FBB values, the stability diagrams of these devices seem noisier and no clean double QD feature is observed. Comparison with these results further supports the selection of EOT-2 for 22FDX® qubits and suggests that our screening process is indeed working as advertised.

## V. Conclusion

Using effective carrier mobility as a figure of merit, we screened six potential processes of reference from GlobalFoundries™ and identified Split F (NMOS, GS-2, EOT-2) and FBB as the best conditions for 22FDX® qubits. Our commercially fabricated qubits formed capacitively- and tunnel-coupled QDs both along the Si channel as well as across it. This is evidenced by the characteristic honeycomb pattern clearly observed in the presented stability diagrams, the behavior of which cannot be attributed to dopants or defects.

Our mobility data as a function of temperature and back bias show that FBB is highly effective at reducing the scattering associated with the front gate, irrespective of EOT and gate stack materials, by inducing back-channel conduction. Moreover, EOT-2 is more effective at shielding carriers from high-$k$ defects, irrespective of applied back bias, and with FBB allows one to achieve the same maximum effective mobility as similar devices with a thicker $SiO_2$ gate oxide [14]. These improvements with FBB however may be limited by intersubband scattering, suggesting that even greater mobility improvements could be seen with FBB of thicker Si channel devices. Overall, this MOSFET screening process represents a viable characterization protocol that can be applied to future qubit nodes.